\documentstyle[11pt,epsf,times]{l-aa}
\def\kmsMpc{km ${\rm s}^{-1}{\rm Mpc}^{-1}$}
\def\hMpc{{{$h^{-1}$}Mpc}}                  
\def\Msun{$h^{-1}{\rm M}_{\odot}$}          
\def\ea{et~al.~}                            
\def\u{\"u}                                 

\def\lesssim{\mathrel{\hbox{\rlap{\hbox{\lower4pt\hbox{$\sim$}}}\hbox{$<$}}}}
\def\gtrsim{\mathrel{\hbox{\rlap{\hbox{\lower4pt\hbox{$\sim$}}}\hbox{$>$}}}}

\newcommand{\Lcdm}{$\Lambda$CDM }
\newcommand{\LCDM}{\bf \Lambda{\rm \bf CDM} }

\begin{document}

\thesaurus{12.03.3; 12.12.1}
\title{Formation of Groups and Clusters of Galaxies}
\author{Alexander Knebe \and Volker M\u ller}
\institute{Astrophysikalisches Institut Potsdam, An der Sternwarte 16,
  D-14482 Potsdam, Germany }
\date{Received date; accepted date}
\maketitle

\begin{abstract}
The formation, inner properties, and spatial distribution of galaxy groups 
and clusters are closely related to the background cosmological model. We use 
numerical simulations of variants of the CDM~model with different cosmological 
parameters and distinguish relaxed objects from recent mergers using the
degree of virialisation. Mergers occur mostly in deep potential wells and 
mark the most important structure formation processes. As consequence, 
the autocorrelation function of merged halos has a higher amplitude and 
is steeper than that of the virialized clusters. This can be directly
connected to the same result found from the observation of luminous 
infrared galaxies; they are as well more strongly correlated  and can
primarily be identified with ongoing merging events.
\end{abstract}

\section{Introduction}
Gravitational instability is commonly accepted as the basic mechanism for 
structure formation on large scales. Combined with the CDM model it leads 
to the picture of hierarchical clustering with wide support from deep galaxy 
and cluster observations. During the recent phase of cosmic evolution 
groups and clusters of galaxies condense from large scale density 
enhancements, and they grow by accretion and merger processes of the 
environmental cosmic matter. We expect traces of such recent formation 
processes still in presently observed objects, such as deviations from 
virial equilibrium, eccentric shapes, and substructures. X-ray 
observations of clusters show a large variety of morphologies which 
hopefully could discriminate between different cosmologies (Mohr \ea 1995).
In low $\Omega_0$ universes structure formation ceases at earlier times 
compared to cosmological models with a higher value for the density parameter.
In the contrary, recent simulations show a remarkable similarity of internal 
structures as the average density profile when they are compared at rescaled 
radii normalized to a typical scale as the virial radius (Jing \ea 1995, 
Thomas \ea 1997). For these analysis they concentrate on a sample of 
isolated clusters in high resolution cosmological simulations which 
emphasizes the relaxed structure found. In the contrary, we intend to 
concentrate on the appearance and the clustering of a sample of simulated 
clusters showing evidence of perturbations and interaction. We discuss 
cluster formation using dissipationless $N$-body simulations of variants 
of the CDM~model with different cosmological parameters $\Omega_0$, 
$\Lambda_0$, and ${\rm H}_0=100 h$~\kmsMpc, where $h$~is the Hubble parameter 
in units of 100~\kmsMpc. Our goal is to interpret the simulated data of the 
structure and spatial distribution of groups and clusters which can 
discriminate between the models, and which can also help to get an  
understanding of the physics underlying the formation of these objects.

One of the critical questions in $N$-body simulations is the identification 
of groups and clusters of galaxies. There are several
methods such as the standard friends-of-friends algorithm (FOF) introduced
by Davis~\ea~(1985), the spherical overdensity method described
by Lacey~\&~Cole~(1994) or more refined algorithm just as the DENMAX
procedure (Gelb \& Bertschinger 1994) or the hierarchical FOF group finding  
(Klypin \ea 1997). But it is by no means clear which of these methods
best fits the observational procedures for identifying groups and clusters of
galaxies. We decided to use the standard FOF group finder because this
algorithm picks up particle groups without preferring any special geometry.
Additionally we adopted the virial theorem: particle groups that do not 
obey this relation are investigated in detail. Visual inspection of such 
objects often shows the FOF group finder's
tendency to link separate groups by small bridges, and groupings of particles 
with strongly discordant velocities. Quantifying their number and specifics 
helps in understanding some of the most impressive objects in the large scale
matter distribution of the universe. We should remark that among the 
unvirialized groups, there exists a number of low mass 
badly resolved objects which
are not of interest in this study. 

The outline of the paper is as follows. First
we specify the $N$-body simulations. Next we discuss the cluster halos and 
their properties. In Section 4 
we study the spatial clustering of the different groups. 
We conclude with a discussion of our main results. 

\section{Cosmological $N$-body simulations}
We base our analysis on four cosmological models all with an amplitude as
measured by the 4-year COBE experiment, cf. Bunn \& White (1997). First 
we take the standard CDM-model, with critical mass density $\Omega_0 = 1$
and a dimensionless Hubble constant $h=0.5$, which is used as a reference in 
spite of the difficulty in reproducing both large and small scale 
clustering. More realistic variants of the CDM model have a lower matter 
content, as $\Lambda$CDM and OCDM1, which both have $\Omega_0 = 0.3$, 
and $h = 0.7$. For these models the shape parameter of the power spectrum 
$\Gamma \equiv \Omega_0 h = 0.21$ better fits the constraints from 
galaxy and cluster clustering, (e.g. Peacock \& Dodds 1996, Einasto \ea 1998). 
The models were normalized to the full 4-year COBE signal using the 
Boltzmann code CMBFAST developed by Seljak \& Zaldarriaga (1996) and assuming 
a baryon content of $\Omega_b h^2 = 0.0125$ suggested by big bang
nucleosynthesis. The mass variance of the linear input spectrum at a scale
of $8 h^{-1}$Mpc is given by $\sigma_8$. Finally we used a less extreme open 
model OCDM2 with $\Omega_0 = 0.5$ which promises realistic large scale
galaxy clustering and has the same mass variance at $8 h^{-1}$Mpc as galaxies.
The information about the models can be found in Table~\ref{parameter}.

\begin{table}
\caption{Physical properties of the numerical simulations. The box size~$L$ 
         is given in~$h^{-1}$Mpc, the particle mass~$m_p$ in units of 
         $10^{11}~$\Msun.}
\label{parameter}
 \begin{tabular}{|l||c|c|c|c|c|c|c|} \hline
             & $\Omega_0$ & $\Omega_{\Lambda ,0}$ & $h$ 
             & $\sigma_8$   & $L$  & $m_p$ \\ \hline \hline
 {\bf SCDM}  & 1.0 & 0.0 & 0.5 & 1.18 & 200 & 11.0 \\ \hline
 {$\LCDM$}   & 0.3 & 0.7 & 0.7 & 1.00 & 280 & 8.7  \\ \hline
 {\bf OCDM1} & 0.3 & 0.0 & 0.7 & 0.48 & 280 & 8.7  \\ \hline
 {\bf OCDM2} & 0.5 & 0.0 & 0.7 & 0.96 & 280 & 15.0 \\ \hline
 \end{tabular}
\end{table}

\begin{figure}[ht]
\unitlength1cm
\begin{picture}(9.5,5.)            
\put(-0.3,-2)                      
        {\epsfxsize=9.5cm \epsfbox{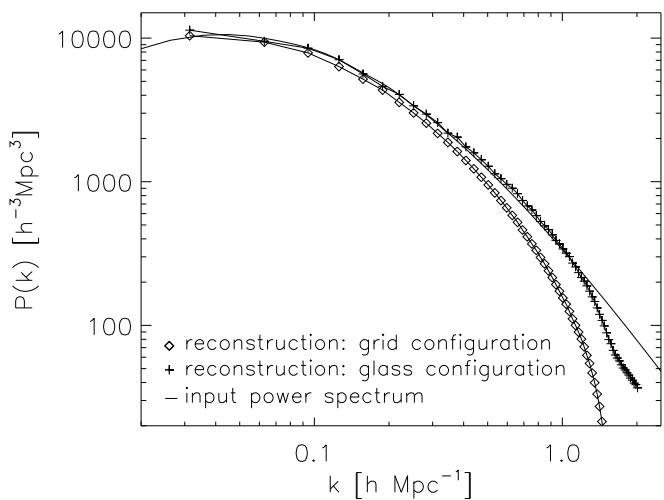}}   
\end{picture}
\vspace*{0.8cm}
\caption{Power spectrum reconstructed from the initial configuration (glass 
         and grid starting points) for the SCDM~model in comparison 
         to the theoretical input spectrum.}
\label{power}
\end{figure}

We employ a `glass' initial particle distribution (White 1996, cf. also 
Couchman \ea 1995) and use the Zeldovich approximation for initial 
displacements and initial velocities according to the specified power 
spectrum. In Fig.~\ref{power} we illustrate the improvement in the 
reproduction of the spectrum in using the irregular versus a grid initial 
particle distribution. The evolution of the gravitating system of dark 
matter particles is simulated with the adaptive ${\rm P}^3$M~code 
(Couchman~1991), adapted to arbitrary background cosmological models. 
The simulations are started at redshift~$z = 25$ and run with 
$128^3$~particles on a $128^3$~grid with a force resolution of 
50~$h^{-1}$kpc~(comoving). 

\section{Cluster Halos and their Properties}
Groups and clusters of galaxies are identified with halos of dark 
matter particles picked out using a standard friends-of-friends algorithm 
which collects particles in groups which have neighbours with separations 
$r_{ll}$ smaller than $ll$~($l$inking~$l$ength) times the mean 
inter particle spacing~$\Delta x = \overline{\rho}^{ \ \frac{1}{3}}$

\begin{equation}\label{lldef}
r_{ll} \leq ll \cdot \Delta x.
\end{equation}

This group finding algorithm does not assume a special geometry for the 
identified objects, but it is based only on particle positions. It picks 
out particle groups for which the density in the outer parts corresponds 
to the value found from the theory of a spherical top hat collapse.

\begin{table}
\caption{Linking lengths~$ll$ for the different models 
         used with the friends-of-friends algorithm
         and the percentage of identified unvirialized 
         particle groups. The mass cut is measured in $10^{14}$~\Msun 
         which corresponds to~25 particles.}
\label{linklength}
 \begin{tabular}{|l||c|c|c|c|} \hline
             & \ $ll$  \ 
             & \ $\delta_{\rm  TH}$ \ 
             & unvirialized halos   
             & mass cut   \\ \hline \hline
 {\bf SCDM}  & \ \ 0.20 \ \   & 178  & 7 \%  & 0.28  \\ \hline
 {$\LCDM$}   & 0.16           & 334  & 4 \%  & 0.22  \\ \hline
 {\bf OCDM1} & 0.15           & 402  & 1 \%  & 0.22  \\ \hline
 {\bf OCDM2} & 0.17           & 278  & 3 \%  & 0.38  \\ \hline
 \end{tabular}
\end{table}

The density 
contrast~$\delta_{\rm TH} = (\rho - \overline{\rho})/\overline{\rho}$ for 
a spherically symmetric (top-hat) collapse in an Einstein-de-Sitter 
universe is $\delta_{\rm TH}\approx 178$. For an isothermal density profile 
$\rho(r) \propto 1/r^2$ this value corresponds to the linking 
length~$ll = 0.2$ (Lacey \& Cole 1994). The linking length for 
gravitationally bound objects in open and $\Lambda$-universes is given by 
\begin{equation}\label{ll}
ll (\delta_{\rm  TH}) = 0.2 \ \displaystyle 
              \left( 
              \frac{178}{\delta_{\rm  TH} (\Omega_0, \Omega_{\Lambda, 0})} 
              \right)^{1/3}.
\end{equation}
\noindent
The values of~$ll$ listed in Table~\ref{linklength} are based on eq.~\ref{ll} 
where the values of $\delta_{\rm  TH} (\Omega_0, \Omega_{\Lambda, 0})$ are 
calculated from the formula found in Katayama \& Suto (1996).

However, it is not clear whether the FOF groups really lead to 
gravitationally bound and virialized groups. The algorithm might link 
two (or more) smaller groups connected by a tenuous bridge
of particles. The group members could also be random encounters, particles 
or particle groups that fly by at the moment of identification, or what
we will see later, are ongoing merging events. 
To check for such effects we explicitly tested the virial 
theorem~$|E_{\rm pot}| = 2 \ E_{\rm kin}$ for each individual halo. This seems 
to be a reasonable procedure since virialisation is a very fast process
in nature, and since unvirialized objects may be identified with particular
processes within the hierarchical structure formation scenario.

Following Spitzer~(1969), the potential energy can be approximated by 
\begin{equation}\label{spitzer}
  |E_{\rm pot,approx}| \ \approx \ 0.4 \ \ G \ \ \displaystyle \frac{M^2}{r_h},
\end{equation}
\noindent where~$r_h$ is the half-mass radius of the system. 
For the smaller halos the potential energy calculated using
eq.~\ref{spitzer} can be compared with the direct summation
\begin{equation}
   |E_{\rm pot,direct}| = 
  \sum_{i=1}^{N} \sum_{j=i+1}^{N} G \displaystyle \frac{m_i m_j}{|r_i - r_j|}
\end{equation}
\noindent This shows that the factor 0.4 is a good choice for most of the 
clusters.
\begin{equation}
  |E_{\rm pot,approx}| \ \lesssim \ |E_{\rm pot,direct}|
\end{equation}
\noindent Therefore, the approximation given by eq.~\ref{spitzer} is used 
throughout the further analysis.

\begin{figure}[ht]
\unitlength1cm
\begin{picture}(9.5,5.)             
\put(-0.5,-1.8)                     
        {\epsfxsize=9.5cm \epsfbox{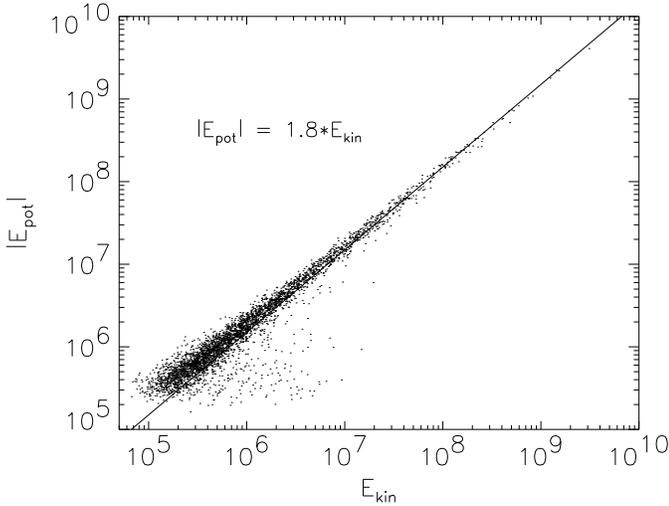}}   
\end{picture}
\vspace*{0.8cm}
\caption{Virial theorem for clusters in the $\Lambda$CDM~model (units are
arbitrary).}
\label{virial}
\end{figure}

The kinetic energy of an identified object is calculated
by summing over all particle velocities with respect to
the motion of the cluster centre
\begin{equation}
 E_{\rm kin} = \frac{1}{2} 
               \sum\limits_{i=1}^{N} 
                m_i (\vec{v_i}-\vec{v_{\rm centre}})^2.
\end{equation}
\noindent
Since all particles have the same mass~$m \equiv m_i$ the kinetic
energy can be connected to the velocity dispersion, 
\begin{equation}
  \sigma_v^2 = \frac{2}{N m} E_{\rm kin}.
\end{equation}

Fig.~\ref{virial} shows the virial relation as a scatter plot 
for all particle groups identified in the \Lcdm model at a redshift of $z = 0$.
The `lower' virial relation shown by the straight line, 
$|E_{\rm pot}| \approx 1.8 \ E_{\rm kin}$, 
can be understood as the influence of an outer pressure of radially infalling 
particles into the halos (Cole~\&~Lacey~1996). The bend of the
virial relation at low energies towards the theoretical expectation 
proves that the surface influence becomes less important for groups of small 
particle numbers, i.e. their stage is less influenced by a steady accretion 
rate than the high mass halos. An opposite effect concerns a much smaller number of 
groups with velocity dispersions which are too high compared to their 
potential energy. These clusters are considered unvirialized and are treated 
separately in the following analysis, and are investigated in detail. There are
many more unvirialized FOF-groups of low and intermediate richness. 
In Table~\ref{linklength} we give the percentage of
unvirialized objects among all FOF clusters for the different cosmological
models. It becomes obvious that the percentage of unvirialized groups is 
highest in SCDM and lowest in OCDM1. This 
reflects to early formation of groups in low $\Omega_0$ models, but also the 
\Lcdm  model has a higher percentage of unvirialized groups in 
comparison to OCDM1.

Now we investigate in more detail the nature of the halos that 
are not gravitationally bound. 

\subsection{Velocity Dispersion -- Mass}
First we discuss the correlation of velocity dispersion 
and the mass of virialized and unbound groups. We expect a 
tight correlation of both quantities for virialized objects if we 
assume an isothermal sphere for the halo and a cutoff at constant
density, which can easily be taken to be the mean cosmic density of 
the halos at the time when they form or suffer their last big merger.
Since most halos form quite recently this cutoff is almost constant, 
and we get  

\begin{equation}
 \sigma_{\rm v} = V_{\rm c}/\sqrt{2} \propto M^{\frac{1}{3}}.
\end{equation}
 
\begin{figure}[ht]
\unitlength1cm
\begin{picture}(9.5,5.)            
\put(-0.3,-1.8)                    
        {\epsfxsize=9.5cm \epsfbox{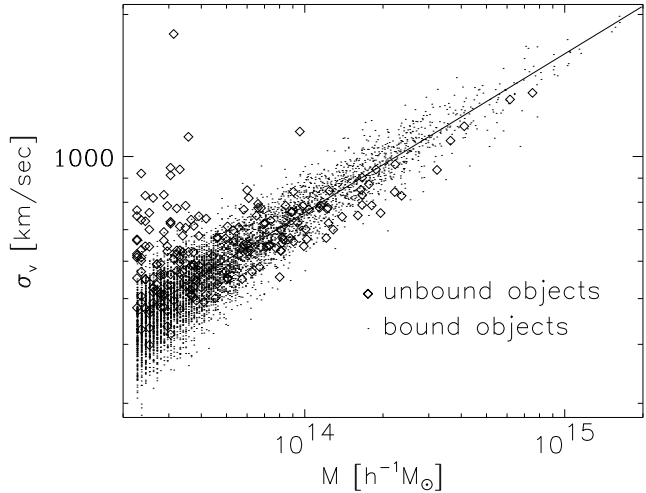}}   
\end{picture}
\vspace*{0.8cm}
\caption{Relation between velocity dispersion and the mass for the 
\Lcdm~model. The virialized halos are marked with points and the objects that 
are not in virial equilibrium are marked using diamonds. The solid line 
is a fit to the scaling 
relation~$\sigma_{\rm v} \propto M^{\frac{1}{3}}$.}
\label{mvdisp}
\end{figure}

As can be seen from Fig.~\ref{mvdisp} this is represented very well in our 
simulations with a large scatter for the light clusters which was also found 
by Cole~\&~Lacey~(1996); we have fitted the data for the virialized particle 
groups to the scaling relation $\sigma_{\rm v}\propto M^{\alpha}$ where 
$\alpha$ was found to be exactly 1/3.
Fig.~\ref{mvdisp} also suggests that the unvirialized objects 
mostly lie at the low mass end of the distribution with high 
velocity dispersions. 
Cole~\&~Lacey~(1996) argue that the tail of groups with 
$\sigma_v > V_{\rm c}/\sqrt{2}$ likely represent objects that belong to 
larger virialized structures. But if 
we repeat our analysis using a larger linking 
length $ll$ we derive similar results: even for FOF groups identified 
with a linking length corresponding to an overdensity 
$\delta \ll \delta_{\rm  TH}$ we find unbound groups in the 
tail $\sigma_v > V_{\rm c}/\sqrt{2}$ in contradiction to the claim of 
Cole~\&~Lacey~(1996). It should be mentioned that at the low mass end the FOF 
groups are not well resolved because these objects consist only of
few particles.
More interesting than the low mass groups are the small number of
halos of all masses, lying beneath the virial relation $\sigma_v = V_{\rm c}/\sqrt{2}$. 
They represent groups in the process of merging which are unbound because
they are too extended. In most cases they represent halos with more then
one centre which are connected by slight tidal bridges and similar structures.
These systems are not very frequent, i.e. they do not survive for a long
time, but they mark most interesting places in the simulation box. And it
can be expected that they are sites for active structure formation processes
in nature.

The results for the different cosmological models introduced in 
Table~\ref{parameter} are comparable and therefore we only a plot 
for the \Lcdm model is presented.

\subsection{Eigenvalues of the Inertia Tensor}
\begin{figure}[ht]
\unitlength1cm
\begin{picture}(9.5,5.)            
\put(-0.5,-1.8)                    
         {\epsfxsize=9.5cm \epsfbox{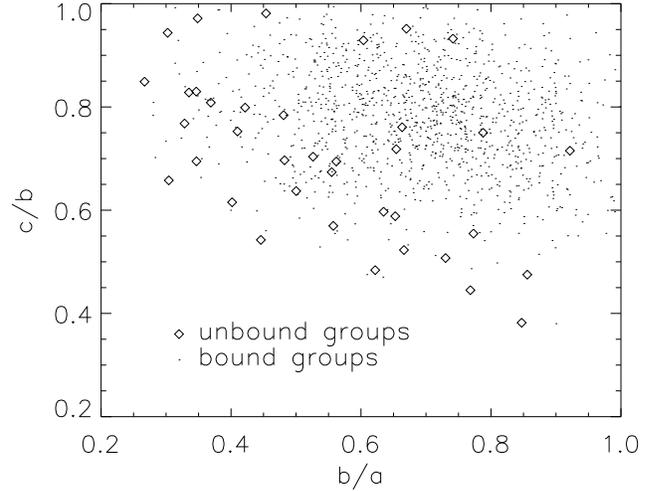}}   
\end{picture}
\vspace*{0.8cm}
\caption{Axis ratio of virialized (dots) and unvirialized groups (diamonds) 
         for the $\Lambda$CDM-simulation at $z=0$.}
\label{axis}
\end{figure}

The next point we are going to investigate is the shape and the axis ratios 
of the identified galaxy clusters. To this aim we calculate the eigenvalues 
of the inertia tensor for each FOF group, which are given in the order 
$a > b > c$. In Fig.~\ref{axis} we show a scatter plot of the ratios of 
the eigenvalues for the $\Lambda$CDM-simulation at $z=0$. In this plot, 
spherical groups are situated in the upper right corner, oblate clusters 
in the right and prolate clusters in the upper part. It is well known 
(Dubinsky 1992, Warren \ea 1992, Cole~\&~Lacey~1996) that hierarchical 
clustering leads to triaxial ellipsoids with a typical axis ratio of 
1:0.7:0.5 which is well represented by our simulations. 

The unvirialized halos inhabit more the lower left part of the diagram, 
again characterizing the soft merging, i.e. the elongation of the groups 
due to tidal interaction of the progenitors which marks the direction of 
the encounters, and in some cases to an elongation of the clusters in a 
second direction due to non-central encounters. Typically unvirialized 
clusters have an axis ratio of 1:0.4:0.3, i.e. they are more oblate and in 
general deviate stronger from sphericity than virialized clusters. 

These findings again to not depend strongly on the the cosmological
background model. We get similar axis ratios when looking at the
SCDM or the OCDM2 model.

\subsection{Impact Parameters}
Let us now discuss how the impact parameter and the relative motion 
between mergers influences the virialisation of the merger products. 
To this aim we consider merging events occurring from redshift $z=0.1$ to
$z=0.0$ between particle groups containing more than 25 particles, and we 
specialize on halos for which the  mass ratio of the two most massive 
progenitors is greater than 1/3. 
The results are listed in Table~\ref{impact} where we have neglected the
analysis of the OCDM1 model; this model is outstanding because of our
normalisation. It does not produce enough galaxy clusters so that
the examination of merging events will not provide reasonable results as
can be seen by the very low percentage of unvirialized objects in
Table~\ref{linklength}.

\begin{table}
\caption{Mean values for the impact parameter (in $h^{-1}$ Mpc) 
         separated for virialized and unvirialized objects.}
\label{impact}
 \begin{tabular}{|l||c|c|}                                      \hline
                        &    virialized objects
                        &  unvirialized objects              \\ \hline \hline
  {\rm \bf SCDM}        &  \raisebox{0.32cm}{} 0.92 $\pm$ 0.60 & 1.19 $\pm$ 0.59  \\ \hline
  {$\LCDM$}             &  \raisebox{0.32cm}{} 0.91 $\pm$ 0.45 & 1.18 $\pm$ 0.48  \\ \hline
  {\rm \bf OCDM2}       &  \raisebox{0.32cm}{} 0.91 $\pm$ 0.48 & 1.26 $\pm$ 0.50  \\ \hline
 \end{tabular}
\end{table}

Using the same constraints on the identification of mergers we have 
calculated the mean values for the impact velocities. The results 
can be found in Table~\ref{vimpact} and show that the two most massive
progenitors of clusters that are identified to be not in virial equilibrium
collide with slightly higher velocities compared to the impact velocities
of progenitors leading to virialized objects.

\begin{table}
\caption{Mean values for the impact velocities (in km/s) 
         separated for virialized and unvirialized objects.}
\label{vimpact}
 \begin{tabular}{|l||c|c|}                                      \hline
                        &    virialized objects
                        &  unvirialized objects              \\ \hline \hline
  {\rm \bf SCDM}        &  \raisebox{0.32cm}{} 670 $\pm$ 500 & 900 $\pm$ 450  \\ \hline
  {$\LCDM$}             &  \raisebox{0.32cm}{} 560 $\pm$ 300 & 600 $\pm$ 280  \\ \hline
  {\rm \bf OCDM2}       &  \raisebox{0.32cm}{} 640 $\pm$ 380 & 830 $\pm$ 350  \\ \hline
 \end{tabular}
\end{table}

As we can see from Table~\ref{impact} the mean value of the impact
parameter for non-virialized particle groups is slightly larger than the 
corresponding value for bound objects. This again is a hint for 
a merging event where the two progenitors had not enough time 
to relax to a gravitational bound object, they have just started to conflate.

\subsection{Virialisation}
If the unvirialized objects are ongoing mergers they should lead to
gravitational bound objects after a short phase of relaxation. This is going 
to be checked now. We identify unvirialized particle groups at a 
redshift of $z=0.1 \ (0.2)$ and have a look at the percentage of these 
objects that are found to be in virial equilibrium at a redshift 
of $z=0.0 \ (0.1)$.

\begin{table}
\caption{Percentage of objects that have virialized from redshift
         $z=0.2$ to $z=0.1$ and $z=0.1$ to $z=0.0$. Again a masscut 
         of 25 particles has been used.}
\label{virialize}
 \begin{tabular}{|l||c|c|}                                      \hline
                        & $z=0.2 \rightarrow z=0.1$
                        & $z=0.1 \rightarrow z=0.0$   \\ \hline \hline
  {\rm \bf SCDM}        & (74 $\pm$ 4) \%  & (63 $\pm$ 4) \%  \\ \hline
  {$\LCDM$}             & (46 $\pm$ 6) \%  & (90 $\pm$ 5) \%  \\ \hline
  {\rm \bf OCDM2}       & (83 $\pm$ 7) \%  & (84 $\pm$ 7) \%  \\ \hline
 \end{tabular}
\end{table}

We can see that most of the unbound particle groups virialize between the 
considered redshifts, for the \Lcdm model nearly all objects have
settled to virial equilibrium within the redshift interval
$z=0.1 \rightarrow z=0.0$. This shows that the time during which 
the cluster is unbound is very short, 
virialisation is a fast process. The low percentage for the \Lcdm model
at the early redshift can be explained, because we have found that
most of the halos had heavy interaction with the surroundings
from redshift $z=0.2$ to $z=0.1$, and so they remained in their non virial 
state. Additionally some of the objects identified at redshift $z=0.2$ 
do not exist anymore at the later redshift: they got disrupted by 
tidal forces from nearby objects. 

\begin{figure}[ht]
\unitlength1cm
\begin{picture}(8.5,11.5)          
\put(0.1,-1.8)                    
        {\epsfxsize=8.5cm \epsfbox{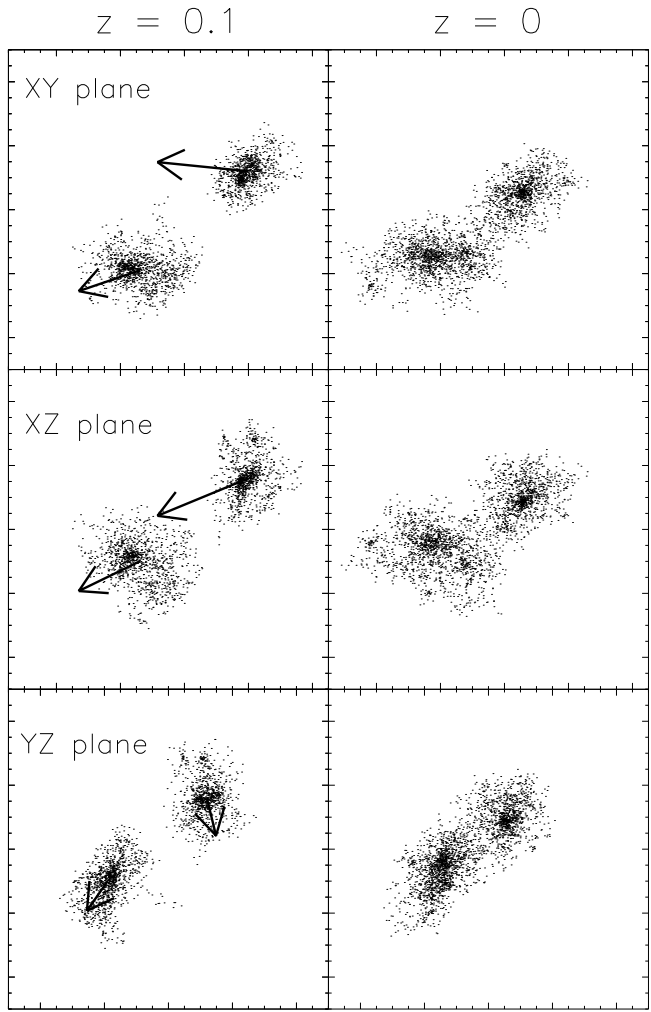}}   
\end{picture}
\vspace*{1.5cm}
\caption{Projections of a cube of $(10 h^{-1} {\rm Mpc})^3$ of the SCDM 
         simulation box at redshifts $z=0.1$ and $z=0$, containing the
         most massive unvirialized cluster. Only the particles in the
         cluster and in the two most massive progenitors are shown which 
         are both virialized objects.}
\label{cluster}
\end{figure}

Fig.~\ref{cluster} shows the most massive unvirialized cluster in the 
SCDM simulation at $z=0$ ($M = 3 \cdot 10^{13}$\Msun) and its two almost 
equal mass progenitors at $z=0.1$. Obviously this is quite a soft merger 
which means that the relative velocity at $z=0.1$ is not much larger
than the inner velocity dispersion. The progenitors perform an almost 
central encounter, during which only a small orbital angular momentum is 
transferred to increase the spin of the resulting halo, 
and there is a quick settling into a new virial equilibrium as the small 
percentage of unvirialized objects proves. The absolute value of the
velocity vector shown in this plot is arbitrary, just the direction and 
the relative lengths are correct to give and impression of how the two 
progenitors collide.

\section{Correlation functions}
One of the basic constraints of cosmological models is the shape
and the amplitude of the two-point correlation function. For many years 
it has been the standard way to describe the clustering of galaxies
and galaxy clusters. The assumption that galaxies (and 
clusters) only form from high-density regions above some threshold 
value~$\delta_c$ leads to a correlation of points exceeding this 
value~$\delta_c$ that is enhanced in comparison to the dark matter 
correlation function (Kaiser 1984). This effect suggests the introduction 
of a biasing parameter $b$,
\begin{equation}
 \xi_{\rm cluster} (r) = b^2 \xi_{\rm mass} (r) .
\end{equation}

\begin{figure}[ht]
\unitlength1cm
\begin{picture}(9.5,5.)           
\put(-0.25,-2)                    
        {\epsfxsize=9.5cm \epsfbox{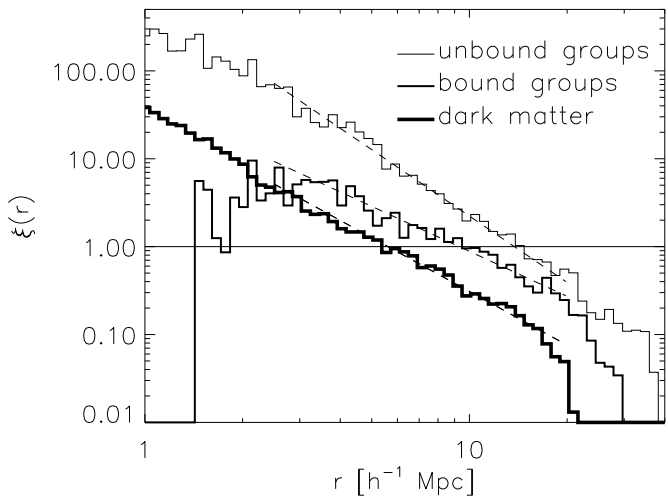}}   
\end{picture}
\vspace*{1.cm}
\caption{Correlation functions for the \Lcdm model. Histograms from below
         (at $r > 3 h^{-1}$Mpc) denote dark matter particles, virialized
         clusters (fixed number density of $n = 10^{-5} h^3 {\rm Mpc}^{-3}$)
         and unvirialized groups. The dashed lines are fits with 
         parameters specified in Table~\ref{xiparam}.}
\label{xi}
\end{figure}

But it is by no means compelling that the bias factor is independent of 
scale as we can see from Table~\ref{xiparam} and Fig.~\ref{xi}.

If we apply this correlation statistics to our simulations we find that
the unvirialized objects are more strongly correlated than the relaxed
systems. In Fig.~\ref{xi} we show this result again only for the 
\Lcdm~model. We find that the correlation between virialized groups 
lies only a factor two over the correlation function of dark matter. 
We fitted standard power laws 

\begin{equation}
\xi(r) = (r_0/r)^{\gamma}
\end{equation}

\noindent which are shown in the Fig.~\ref{xi} as dashed lines. Galaxy 
groups in the \Lcdm model have a correlation length $r_0 \sim 9$ \hMpc \ 
and a slope $\gamma = 1.7$. The correlation function turns negative 
beyond 30 \hMpc. Fits for the other models are given in Table~\ref{xiparam}. 
To get comparable values for the different cosmologies we fixed the 
cluster number density to $n = 10^{-5} h^3 {\rm Mpc}^{-3}$ and used only the 
$N = n V$ most massive clusters for the calculation of the correlation 
function (values for $N$ are given in Table~\ref{xiparam}, too). Obviously 
the amplitude in SCDM and OCDM2 are smaller than in the \Lcdm model, i.e., 
these models have too small power on large scales for describing the 
clustering of groups and clusters of galaxies. 

\begin{table}
\caption{Fit parameter of the correlation functions for dark matter
         (first column), bound clusters (second column) and unbound 
         clusters (third column) for three cosmological models.}
\label{xiparam}
 \begin{tabular}{|l||c|c||c|c|c||c|c|} \hline
                  & \multicolumn{2}{c||}{dark matter}
                  & \multicolumn{3}{c||}{bound groups} 
                  & \multicolumn{2}{c|}{unbound groups}                    \\ \hline 
                  & $r_0$ & $\gamma$ & $r_0$ & $\gamma$ & $N$  &$r_0$  & $\gamma$ \\ \hline \hline
  {\rm \bf SCDM}  &  5.4  &   2.3    &  6.6  &    1.8   & 800  &  9.2  &   2.5    \\ \hline
  {$\LCDM$ }      &  5.6  &   2.0    &  9.3  &    1.7   & 2200 & 13.8  &   2.5    \\ \hline
  {\rm \bf OCDM2} &  4.5  &   2.2    &  7.9  &    2.0   & 2200 & 11.4  &   2.5    \\ \hline
 \end{tabular}
\end{table}

\begin{table}
\caption{Parameter for the correlation functions calculated for 
         different mass cuts for the virialized halos. The mass cut
         is given as the number of particles.}
\label{ximass}
 \begin{tabular}{|c||c|c||c|c||c|c|} \hline
                  & \multicolumn{2}{c||}{\rm \bf SCDM}
                  & \multicolumn{2}{c||}{$\LCDM$} 
                  & \multicolumn{2}{c|} {\rm \bf OCDM2}  \\ \hline 
   mass cut       & $r_0$ & $\gamma$ & $r_0$ & $\gamma$ & $r_0$ & $\gamma$ \\ \hline \hline
  25   & 4.0 & 2.1 & 6.2 & 1.8 & 5.6 & 1.7  \\ \hline
  50   & 4.5 & 1.8 & 8.3 & 1.7 & 6.7 & 1.9  \\ \hline
  100  & 5.1 & 1.7 & 9.1 & 1.7 & 8.4 & 2.1  \\ \hline
 \end{tabular}
\end{table}

In Table~\ref{ximass} we demonstrate the strong dependence of the amplitude 
of the correlation function on the mass cuts. High mass halos mark high 
density maxima in the dark matter distribution which are stronger clustered 
(Kaiser 1984). The slope remains almost unchanged. Due to the smallness of 
the simulation box it is difficult to compare with the correlation function 
of rich clusters of galaxies which are identified with halos of mass larger 
than $2 \cdot 10^{14}$ \Msun. It is interesting that the bound halos in the 
SCDM model have the lowest amplitude among all models. This is connected 
with the high evolution stage of this model where very big halos form which 
include most of the matter in the simulation box, and on the other hand with 
the small power of SCDM at large with respect to small scales, this being 
related to the high value of $\Gamma=0.5$. Observations lead to a preferred 
value of $\Gamma=0.25\pm 0.05$  (Peacock \& Dodds 1996).

\begin{figure}[ht]
\unitlength1cm
\begin{picture}(9.5,5.)           
\put(-0.26,-3)                    
        {\epsfxsize=9.5cm \epsfbox{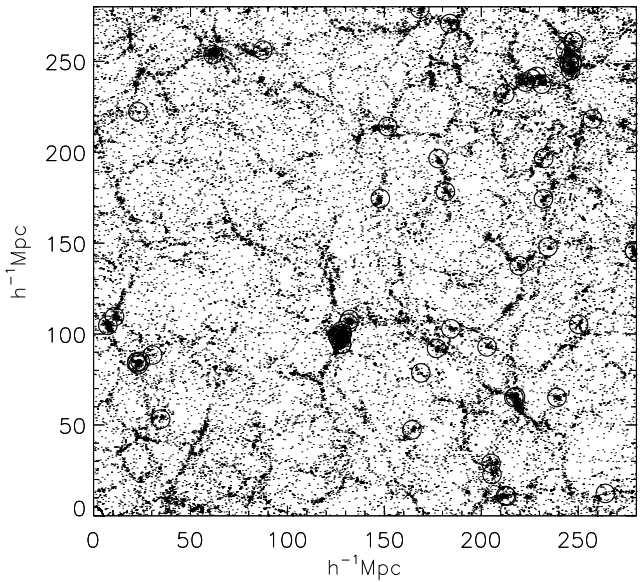}}   
\end{picture}
\vspace*{2.3cm}
\caption{Slice through the \Lcdm simulation. The circles indicate the
         positions where unvirialized objects have been found.}
\label{slice}
\end{figure}

But most interesting is the correlation function of the unvirialized 
particle groups.
They are much more strongly clustered than virialized objects, and the slope 
is very steep, $\gamma = 2.5$, independent of the cosmological model, cf. 
Table~\ref{xiparam}. This verifies that mergers are occurring at 
particular places in the universe, and that these processes are highly 
correlated. This can be checked in Fig.~\ref{slice}. We have plotted a 
slice through the \Lcdm simulation and marked the places where unvirialized
objects have been identified with circles. These objects are found preferably
in high density regions whereas virialized halos trace the overall large 
scale structure of the universe.

Nowadays it seems clear that strong interactions and mergers are the 
trigger for producing the most luminous infrared galaxies. So an enhanced 
clustering 
has been obtained for luminous infrared galaxies (Bouchet \ea 1993). 
More luminous IRAS galaxies, basically mergers, are more strongly
correlated, and ultraluminous mergers have strongest correlation strength
(Gao 1993).
These luminous infrared galaxies may represent an important stage in the 
formation of QSO's and 
a primary stage in the formation of elliptical galaxies cores. And as we
have found, these objects form at special places in the universe where
the most massive galaxy clusters are found.

\section{Conclusions}
We have collected a number of reasons for looking at virialisation as a 
reasonable criterion for identifying groups and clusters of galaxies in
numerical simulations. The virial theorem well described an 
overwhelming majority of the halos in our simulations. A slight bend in 
the virial relation can be ascribed to the effective pressure of the 
permanent spherical accretion stream in the halos. 

Special attention has been devoted to the unvirialized halos. We have 
shown that they are characterized by quite recent soft mergers which 
lead to more anisotropic halos and are strongly correlated over scales 
of up to 40 \hMpc. This is confirmed by the result obtained for luminous 
infrared galaxies. In our simulations we find a very fast virialisation 
of the halos which leads to the small percentage of such objects at any 
given time. But the virialisation itself was not a basic goal of our 
simulations. It must be checked whether this time depends on the 
relatively low mass resolution in our big simulation boxes. The SCDM model 
is found to have the largest evolution rate of the halo abundance to $z=0$, 
and has the largest percentage of unvirialized objects. On the contrary, 
the more realistic \Lcdm and OCDM2 models have a quite small number of 
recent mergers. 

The stronger anisotropy of the merger products leads to typical triaxial
ellipsoids with axis ratios of 1:0.4:0.3 which represents an measurable effect 
in large cluster samples. The merging processes are 
characterized by basically central encounters. Not much angular momentum 
is transferred to the merger product which leads to a self-similar 
growth of the rotation of the halos, (e.g. Knebe 1998). 

The large differences in the correlation functions of the models makes it
worthwhile to further compare differences in the large-scale matter 
distributions of the different models.

\begin{acknowledgements}
We are grateful to H. Couchman for making his numerical code public. Also 
we acknowledge the use of CMBFAST provided by U. Seljak and M. Zaldarriaga. 
We are glad about useful and encouraging discussions 
with J. Colberg. S. Gottl\"ober and D. Woods read carefully the manuscript 
and made valuable comments.
\end{acknowledgements}



\begin{thebibliography}{}
\bibitem[1993]{bouchet}
{Bouchet F.R., Strauss M.A., Davis M., Fisher K.B., Yahil A., Huchra J.P.,
 ApJ {\bf 417}, 36 (1993)}

\bibitem[1997]{bunn} 
{Bunn E.F., White M., ApJ {\bf 480}, 6 (1997)}

\bibitem[1996]{cole}
{Cole S., Lacey C., MNRAS {\bf 281}, 716 (1996)}

\bibitem[1991]{couchman1}
{Couchman H.M.P., ApJ {\bf 368}, L23 (1991)}

\bibitem[1991]{couchman2}
{Couchman H.M.P., Thomas P.A., Pearce F.R., ApJ {\bf 352}, 797 (1995)}

\bibitem[1985]{davis}
{Davis M., Efstathiou G., Frenk C. S., White S.D.M., ApJ {\bf 292}, 371 (1985)}

\bibitem[1992]{dubinsky} {Dubinski J., ApJ {\bf 401}, 441 (1992)}

\bibitem[1998]{einasto}
{Einasto J., Einasto M., Tago E., Starobinsky A.A., Atrio-Baradela F., 
M\"uller V., Knebe A., Frisch P., Cen R., Andernach H., ApJ submitted (1997)}

\bibitem[1993]{gao}
{Gao, Y., in: The Evolution of Galaxies and
Their Environment, eds. Hollenbach, D., Thronson, H., Shull, J.M.,
NASA Conf. Pub. 3190, Ames: NASA 1993, 54}

\bibitem[1994]{gelb}
{Gelb J., Bertschinger E., ApJ {\bf 436}, 467 (1994)}

\bibitem[1995]{jing}
{Jing, Y.P., Mo, H.J., B\"orner, G., Fang, L.Z., MNRAS {\bf 276}, 417 (1995)}

\bibitem[1984]{kaiser} 
{Kaiser N., ApJ Lett. {\bf 284}, L9 (1984)}

\bibitem[1996]{katayama}
{Katayama T., Suto Y., ApJ {\bf 469}, 480 (1996)}

\bibitem[1998]{knebe} 
{Knebe A., in: Large Scale Structures: Tracks and 
Traces. Proceedings of the 12th Potsdam Cosmology Workshop, 
ed. V. M\"uller \ea, World Scientific 1998, p. 175}

\bibitem[1997]{kravtsov}
{Klypin A., Gottl\"ober S., Kravtsov A.V., (astro-ph/9708191)}

\bibitem[1993]{lacey1}
{Lacey C., Cole S., MNRAS {\bf 262}, 627 (1993)}

\bibitem[1994]{lacey2}
{Lacey C., Cole S., MNRAS {\bf 271}, 676 (1994)}

\bibitem[1995]{mohr}
{Mohr J.J., Evrard, A.E., Faricant, D. G., Geller, M.J., 
ApJ {\bf 447}, 8 (1995)}

\bibitem[1996]{peacock} 
{Peacock J. A., Dodds S.J., MNRAS {\bf 280}, 1020 (1996)}

\bibitem[1996]{seljak} 
{Seljak U., Zaldarriaga M., ApJ {\bf 469}, 437 (1996)}

\bibitem[1969]{spitzer} 
{Spitzer L., ApJL {\bf 158}, 139 (1969)}

\bibitem[1997]{thomas} 
{Thomas P.A., Colberg J.M., Couchman H.M.P., Efstathiou G.P., Frenk C.S.,
Jenkins A.R., Nelson A.H., Hutchings R.M., Peacock J.A., Pearce F.R., 
White S.D.M. MNRAS submitted (astro-ph/9707018)}

\bibitem[1992]{warren}
{Warren M.S., Quinn P.J., Salmon J.K., Zurek W.H., ApJ {\bf 399}, 405 (1992)}

\bibitem[1993]{white}
{White S.D.M., in: Cosmology and Large-Scale Structure, eds. Schaeffer R., 
Silk J., Spiro M., Zinn-Justin J., Elsevier 1996, p. 349}

\end{thebibliography}
\end{document}